\documentclass[sigconf]{acmart}

\usepackage{tikz}
\usetikzlibrary{arrows,calc,chains,shapes,decorations.pathreplacing}
\usetikzlibrary{shapes.geometric,shapes.arrows,decorations.pathmorphing}
\usetikzlibrary{matrix,scopes,positioning,fit}
\usepackage{bm}
\usepackage{booktabs}
\usepackage{pgfplots}
\usepackage{wrapfig}
\usepackage{algorithmic}
\usepackage{graphicx}
\usepackage{textcomp}
\usepackage{xcolor}
\usepackage{balance}
\usepackage{adjustbox}
\usepackage{natbib}
\usepackage{scalerel}
\usepackage[justification=justified,skip=3pt]{caption}
\usepackage{multirow}
\renewcommand{\arraystretch}{1}
\AtBeginDocument{%
  \providecommand\BibTeX{{%
    \normalfont B\kern-0.5em{\scshape i\kern-0.25em b}\kern-0.8em\TeX}}}

\newcommand\blfootnote[1]{%
  \begingroup
  \renewcommand\thefootnote{}\footnote{#1}%
  \addtocounter{footnote}{-1}%
  \endgroup
}

\copyrightyear{2021}
\acmYear{2021} 
\setcopyright{iw3c2w3}
\acmConference[WWW '21]{Proceedings of the Web Conference 2021}{April 19--23, 2021}{Ljubljana, Slovenia} 
\acmBooktitle{Proceedings of the Web Conference 2021 (WWW '21), April 19--23, 2021, Ljubljana, Slovenia}
\acmPrice{}
\acmDOI{10.1145/3442381.3449864}
\acmISBN{978-1-4503-8312-7/21/04}

\begin{document}

\title{Variation Control and Evaluation for Generative Slate Recommendations}


\author{Shuchang Liu$^{1*}$, Fei Sun$^{2\dagger}$, Yingqiang Ge$^1$, Changhua Pei$^2$, Yongfeng Zhang$^1$}
\affiliation{
\institution{$^1$Rutgers University, New Brunswick, NJ, USA \kern3pc $^2$Alibaba Group, Beijing, China}
{$^1$\{shuchang.syt.liu, yingqiang.ge, yongfeng.zhang\}@rutgers.edu $\quad$ $^2$\{ofey.sf, changhua.pch\}@alibaba-inc.com}
}

\renewcommand{\authors}{Shuchang Liu, Fei Sun, Yingqiang Ge, Changhua Pei, and Yongfeng Zhang}
\renewcommand{\shortauthors}{Shuchang Liu, Fei Sun, Yingqiang Ge, Changhua Pei, and Yongfeng Zhang}

% \author{Shuchang Liu}
% \affiliation{
%   \institution{Rutgers University}
%   \city{New Brunswick, NJ}
%   \country{US}
% }
% \email{shuchang.syt.liu@rutgers.edu}

% \author{Fei Sun}
% \affiliation{
%   \institution{Alibaba Group}
%   \city{Beijing}
%   \country{China}
% }
% \email{ofey.sunfei@gmail.com}

% \author{Yingqiang Ge}
% \affiliation{
%   \institution{Rutgers University}
%   \city{New Brunswick, NJ}
%   \country{US}
% }
% \email{yingqiang .ge@rutgers.edu}

% \author{Changhua Pei}
% \affiliation{
%   \institution{Alibaba Group}
%   \city{Beijing}
%   \country{China}
% }
% \email{changhua.pch@alibaba-inc.com}

% % \author{Amelie Marian}
% % \affiliation{
% %   \institution{Rutgers University}
% % }
% % \email{amelie@rutgers.edu}

% \author{Yongfeng Zhang}
% \affiliation{
%   \institution{Rutgers University}
%   \city{New Brunswick, NJ}
%   \country{US}
% }
% \email{yongfeng.zhang@rutgers.edu}


\begin{abstract}
Slate recommendation generates a list of items as a whole instead of ranking each item individually, so as to better model the intra-list positional biases and item relations. 
In order to deal with the enormous combinatorial space of slates, recent work considers a generative solution so that a slate distribution can be directly modeled.
However, we observe that such approaches---despite their proved effectiveness in computer vision---suffer from a trade-off dilemma in recommender systems:
when focusing on reconstruction, they easily over-fit the data and hardly generate satisfactory recommendations;
on the other hand, when focusing on satisfying the user interests, they get trapped in a few items and fail to cover the item variation in slates. 
In this paper, we propose to enhance the accuracy-based evaluation with slate variation metrics to estimate the stochastic behavior of generative models.
We illustrate that instead of reaching to one of the two undesirable extreme cases in the dilemma, a valid generative solution resides in a narrow ``elbow'' region in between.
And we show that item perturbation can enforce slate variation and mitigate the over-concentration of generated slates, which expand the ``elbow'' performance to an easy-to-find region.
We further propose to separate a pivot selection phase from the generation process so that the model can apply perturbation before generation.
Empirical results show that this simple modification can provide even better variance with the same level of accuracy compared to post-generation perturbation methods.
\blfootnote{$^*$This work was done when Shuchang Liu was an intern at Alibaba.}
\blfootnote{$^{\dagger}$Corresponding author.} 
\end{abstract}




\maketitle

\section{Introduction}
\label{sec: intro}

In most recommender systems, items are naturally exposed to users as a slate, which usually contains a \textit{fixed} number of items, e.g., a 1-by-5 list of recommended items, or a 2-by-2 block that can fit a mobile phone screen.
This leads to the idea of slate recommendation, also known as exact-$k$ recommendation~\cite{swaminathan2017off,Gong:2019:ERV:3292500.3330832}. 
The problem is usually formalized as generating a slate of items such that certain expected user behavior (e.g., the number of clicks) is maximized.
The challenge of this problem is that the number of possible slates is combinatorially large~\cite{viappiani2010optimal}.
For example, for a system with $n$ items, to generate a slate of $k$ items, the possible number of slates will be $O(n^k)$, which is huge given that many recommender systems work on millions or even billions of items.

Traditional ranking-based recommendation models such as learn-ing-to-rank (LTR)~\cite{burges2005learning, cao2007learning, liu2009learning, rendle2012bpr, ge2020learning} first predicts the probability of user engagement on each candidate item, and then selects the top-ranked ones as the recommendation list.
Despite its well-recognized effectiveness and scalability, this ranking and selection process is greedy in essence and neglects the fact that the user behavior on an item may be influenced by 
other (e.g., complementary or competitive) items exposed in the same list~\cite{joachims2005accurately, yue2010beyond}, thus resulting in its sub-optimality.
Furthermore, evidence has shown that one can improve the recommendation performance by taking into account the intra-list item relations in ranking ~\cite{cao2007learning, 47637, yue2010beyond, qin2009global, dokania2014learning, Gong:2019:ERV:3292500.3330832}.

Recently, researchers have explored the possibility of solving this problem by directly generating the slate as a whole to break the limitation of ranking-based approaches. 
Many of the approaches are based on generative models such as Variational Auto-Encoders (VAE) \cite{jiang2018beyond,higgins2017beta}.
However, these generative models are stochastic in nature and their variational behavior may not produce satisfactory slate recommendations.
For example, in the case of VAE-based models, the performance depends on a trade-off coefficient $\beta$~\cite{higgins2017beta}---the larger the $\beta$-value during training, the more the model is focused on encoding variation control against the data reconstruction accuracy.
In terms of slate recommendation, this phenomenon diverges the generative results into one of the three cases:
\begin{itemize}
    \item \textbf{Over-reconstruction}: when $\beta$ is smaller than some lower threshold $\beta^-$, it tends to overfit the slate reconstruction on the training set.
    Though the resulting generated slates have extremely high variance, the model usually fails to generate satisfactory recommendations.
    \item \textbf{Over-concentration}: when $\beta$ is larger than some upper threshold $\beta^+$, the model tends to choose from only a few prototypical slates that achieve satisfactory performance but fails to explore the variety of slates.
    \item \textbf{Elbow case}: when $\beta$ is selected in an appropriate region (i.e., $\beta \in \left[\beta^{-}, \beta^{+}\right]$), it gives intermediate item variety and is able to fulfill certain degree of user interests.
    We show that this transitional region is the most suitable for slate recommendation task.
    Unfortunately, this very setting usually lies in a narrow region (e.g. $\beta^{+}-\beta^{-} \ll 10^{-2}$) while the search space of $\beta$ can be arbitrarily large.
\end{itemize}
We denote this as the {\em Reconstruction-Concentration Dilemma} (RCD) and in this paper we investigate possible solutions that can increase the variety of items under the over-concentration case.
To achieve this, one can simply apply post-generation perturbation to enforce item variety, yet this solution ignores the intra-slate features and significantly downgrades the recommendation accuracy.
With this in mind, we further derive a modification of the original generation process, so that it can perturb before the final generation while reducing the negative effect of the perturbation.
Specifically, when generating a slate, it follows a two-phase procedure: first, a pivot selection model chooses an item for a fixed slate position; then a slate completion model generates the remainder of the slate based on the pivot item along with other constraints.
With this framework, we summarize our contributions as below:

\begin{itemize}
    \item We propose to consider both the slate accuracy metric and the slate variation metric when evaluating models that generate stochastic slates.
    \item We identify the RCD with these metrics and show that the most desirable recommendation performance appears in a narrow ``elbow'' region.
    \item We conduct experiments on real-world datasets and simulation environments to show that enforcing variation can mitigate over-concentration and extend the elbow's performance to a wide range of search space.
    \item We show that the proposed pivot selection phase can provide better control over the slate variation under the over-concentration case of the dilemma.
\end{itemize}

In the following sections, we first list related studies in section \ref{sec: related_work}, then describe how generative slate recommendation is achieved in section \ref{sec: list_cvae}.
Further, we explain how to employ variance metrics as complements of accuracy metrics in section \ref{sec: variance}, and then introduce our slate recommendation framework 
in section \ref{sec: framework}.
We present our experimental results on both real-world datasets and simulation environments in section \ref{sec: experiments} and \ref{sec: results} as the evidence to support our claims.
And finally, we discuss some other possible solutions that may also improve the item variety to bridge the gap between generative methods and recommendation systems.

%%%%%%%%%%%%%%%%%%%%%%%%%%%%%%%%%%%%%%%%%%%%%%%%
%                Related Work                  %
%%%%%%%%%%%%%%%%%%%%%%%%%%%%%%%%%%%%%%%%%%%%%%%%

\section{Related Work}\label{sec: related_work}

There exist several types of generative modeling approaches to recommender systems.
The most studied area is to leverage recurrent neural networks (RNN)~\cite{donkers2017sequential}.
It models the probability of each item conditioned on all previously recommended items $P(d_i|d_{i-1}, \dots, d_1)$ and consecutively make recommendations from $d_1$ to $d_K$.
Modeling in this way means that the recommendation of item $d_i$ does not depend on the items $d_{i+1}, \dots, d_K$ that appear later, which weakens the intra-list relation of the recommendations.
This sub-optimality has already been shown in~\cite{jiang2018beyond}.
Another track of research uses auto-encoder for recommendation \cite{sedhain2015autorec, liang2018variational}, but they model the user history profiles instead of the distribution of slates. A recent line of research adopts reasoning-based recommendation models \cite{shi2020neural,chen2021neural,xian2020cafe}, which models recommendation as a cognition rather than perception task and adopts neural reasoning rather than neural matching models for better recommendation.

In addition to the generative approach represented by \cite{jiang2018beyond}, there are other efforts that aim to deal with slate recommendation using reinforcement learning (RL)~\cite{swaminathan2017off, ie2019reinforcement, ie2019slateq, ge2021towards}. 
Like the early attempts ~\cite{shani2005mdp,taghipour2007usage}, this type of methods mostly targets on exploring how to make use of the long term effects of several consecutive recommendations by transforming the slate and its user reaction as ``states'' in RL.
Though they are suitable for solving the problem of slate recommendation, the essence behind RL and generative methods are mostly complementary, since a generated model can be pretrained and transplanted as the actor in RL frameworks.

We can also consider slate recommendation as a type of list recommendation, but the list size is fixed.
Except for accuracy measures, there are many list-wise metrics that are proved beneficial to both the recommender systems and its customers, including but not limited to coverage~\cite{good1999combining} and intra-list diversity~\cite{ziegler2005improving,zhou2010solving}.
Typically, the solution has to balance between accuracy and diversity, such as Max-Marginal Relevance (MMR)~\cite{carbonell1998use}, relative benefits~\cite{bradley2001improving}, $\alpha$-NDCG~\cite{clarke2008novelty}, and Determinantal Point Process (DPP)~\cite{gartrell2016bayesian}.
But as pointed out by \citeauthor{jiang2018beyond} ~\cite{jiang2018beyond}, it will be unfair to compare these essentially discriminative methods in generative settings, and conversely, it will be unfair for generative methods to compete in traditional LTR settings.
In order to show this deviation, we investigate how much discriminative ranking methods are different from generative methods if compared in the same setting in section~\ref{sec: results}.

A relatively unrelated track that considers slate-wise patterns is to re-rank the items based on the expected user interaction on the candidate slate~\cite{ai2018rerank, bello2018seq2slate, wei2020generator}.
However, the items available for re-ranking are often restricted to the candidates given by some base ranking model. Our problem is about directly generating slate recommendations with no restriction on candidate items, which is essentially a different task.
One should also distinguish slate recommendation with session-based recommendation~\cite{Hidasi2015SessionbasedRW}, which usually consists of user interaction history of arbitrary length, typically with a sequence of sessions, and the major research focus is on the modeling of the user sequential behaviors~\cite{donkers2017sequential, sun2019bert4rec}.

\section{Generative Slate Recommendation}\label{sec: slate_recommendation}

% We adopt similar problem formulation as in \cite{jiang2018beyond}.
The corpus of items is denoted as $\mathcal{D}$, and a slate of size $K$ is defined as an ordered list of items $\bm{s}=(d_1, d_2, \dots, d_K)$, where $d_k{\in}\mathcal{D}$ and positional index $k\in\{1,\dots,K\}$ represents that the item appeared in the $k$-th slot in the slate.
A user's response to a slate $\bm{s}$ is denoted as $\bm{r}=(r_1,r_2,\dots,r_K)$, where $r_k$ is the response on item $d_k$, e.g., $r_k \in\{0,1\}$ represents $d_k$ is clicked or not.
Assume that each slate $\bm{s}$ has corresponding latent unknown features $\bm{z}$ and some known characteristics/constraints $\bm{c}$.
Typically, let $\bm{c}=\mathrm{onehot}(\sum_{k=1}^K r_k)$ so that the user responses are contained in the constraints.
For example, for a slate with 0 click, the corresponding constraint would be $[1,0,0,0,0,0]$, while for a slate with 3 clicks, the constraint would be $[0,0,0,1,0,0]$.
% so that the model indirectly models the $P(\bm{s}|\bm{r})$.
Unlike discriminative ranking methods that model $R(\bm{r}|\bm{s})$, which is the user response for a given slate, the goal of generative slate recommendation models is to learn the distribution of slates with the given constraints:
\begin{equation*}
    P_{\theta}(\bm{s}|\bm{z},\bm{c})
\end{equation*}
where $\bm{z}$ is the latent slate encoding.
% that can directly generate a desired slate $\bm{s}$ given some encoded features $\bm{z}$ and the constraints $\bm{c}$.
An optimal slate $\bm{s}^*$ should maximize the expected number of clicks $\mathbb{E}[\sum_{k=1}^K r_k]$,
% Following the generative methodology, assume that each slate $\bm{s}$ and its user response $\bm{r}$ are sampled from some joint distribution, indicating a possible correlation in between.
so during recommendation, one should provide to the inference model with the maximum number of clicks as constraint $\bm{c}^*= [0,0,0,0,0,1]$ (correspond to the ideal all-clicked response $\bm{r}^*=[1,1,1,1,1]$).
Different from the setting in \cite{jiang2018beyond}, we also allow user features, so the constraint vector $\bm{c}$ in this case will be the concatenation of extracted user embedding and the aforementioned transformed response.
As we will discuss in section \ref{sec: user_vs_nouser}, a more fine-grained constraint vector that involves user is more likely to induce a smooth distribution instead of a disjoint manifold in the encoding space $\bm{z}$.

\subsection{Slate Generation Model}\label{sec: list_cvae}

To find a good generative model $P_{\theta}(\bm{s}|\bm{z},\bm{c})$, a Conditional Variational Auto-Encoder (CVAE) framework learns a set of latent factors $\bm{z}\!\in\!\mathbb{R}^m$ such that $\bm{z}$ can encode sufficient high-level information to reproduce the observed slates with maximum likelihood.
As formulated in \cite{kingma2013auto}, a variational posterior $Q_{\phi}(\bm{z}|\bm{s},\bm{c})$ is used as an approximation to solve the intractable marginal likelihood (which involves integral over latent $\bm{z}$).
The resulting model structure contains an encoder $Q_{\phi}$ that learns to encode the input slate $\bm{s}$ and constraint $\bm{c}$ into a set of variational information 
(e.g., the mean and variance when Gaussian prior is assumed) of each factor of $\bm{z}$, and a decoder $P_{\theta}$, which corresponds to the generative model.
When training the model, one can maximize the variational Evidence Lower Bound (ELBO) of the data likelihood \cite{kingma2013auto}, which is equivalent to minimizing:
\begin{equation}
    \mathcal{L}_{\bm{s}} = \mathbb{E}_{Q_{\phi}(\bm{z}|\bm{s},\bm{c})}\bigl[\log P_{\theta}(\bm{s}|\bm{z},\bm{c})\bigr] - \beta \mathrm{KL}\bigl[Q_{\phi}(\bm{z}|\bm{s},\bm{c}) \Vert P_{\theta}(\bm{z}|\bm{c})\bigr]
\label{eq: cvae_loss}
\end{equation}
where $P_{\theta}(\bm{z}|\bm{c})$ is the conditional prior distribution of $\bm{z}$, KL represents the Kullback-Leibler Divergence (KLD), which restrains the distance measure between two distributions, and $\beta$ is the trade-off coefficient as described in section~\ref{sec: intro}.
The encoder, decoder, and the conditional prior are all modeled by neural networks to capture complex feature interactions.
With the decoder, items of each slate are selected based on the dot product similarity between output embeddings and embeddings of all items in $\mathcal{D}$.
During training, in order to avoid overfitting, the reconstruction loss is calculated by the cross entropy over down-sampled items instead of the entire $\mathcal{D}$.
At inference time, the slate is generated by passing the ideal condition $\bm{c}^*$ into the decoder along with a randomly sampled encoding $\bm{z}$ (e.g., from random Gaussian) based on the variational information provided by the conditional prior.

In the loss Eq.~\eqref{eq: cvae_loss}, we can interpret the KL divergence as how well the learned encoding $\bm{z}$ distribution is regulated by the guiding prior $P_\theta(\bm{z}|\bm{c})$, and the other term reveals how well existing slates are reconstructed.
According to~\cite{higgins2017beta}, manipulating the trade-off parameter $\beta$ will push the model to favor one of the terms over the other.
For example, if we assume isotropic Gaussian as the prior distribution and set larger $\beta$, the factors in the learned $\bm{z}$ space will become more disentangled, and thus more meaningful control over the generation, but with a possible downgrade of reconstruction performance resulting in unrealistic generation.
Despite its feasibility in many other tasks, as we will discuss in section \ref{sec: rcd}, this $\beta$ leads to a reconstruction-concentration trade-off that barely provide satisfactory recommendation results.

%%%%%%%%%%%%%%%%%%%%%%%%%%%%%%%%%%%%%%%%%%%%%%%%
%             Proposed Framework               %
%%%%%%%%%%%%%%%%%%%%%%%%%%%%%%%%%%%%%%%%%%%%%%%%

\subsection{Variance Evaluation of Generated Slates}\label{sec: variance}
Many generative methods (e.g. VAEs and GANs\cite{goodfellow2014generative}) are stochastic in terms of the output, but it is possible that the slate encoding $\bm{z}$ is not obtained through an encoder model so one cannot simply estimate the slate variance based on $\bm{z}$.
Thus, we are interested in evaluation metrics that can estimate the variance of slates for a wide range of stochastic models.

An evident choice is directly using \textbf{item variance} across all possible generated slates.
Since items are typically represented by embedding vectors, let $\bm{x}_1,\dots,\bm{x}_K$ be the vector representations of generated items.
For simplicity, assume conditional independence among factors of $\bm{x}$, then the item variance can be calculated as the variance of each factor and be approximated by sampling: 
\begin{equation}
    \begin{aligned}
    \textrm{Cov}(\bm{x}) & = \mathbb{E}_{\bm{s}\sim P_\theta}\left[\frac{1}{K}\sum_{i=1}^K\bigl\|\bm{x}_i^{(\bm{s})} - \bm{\mu}\bigr\|^2\right] \\ 
    & = \lim_{N\rightarrow\infty}\frac{1}{NK}\sum_{j=1}^{N}\sum_{i=1}^K\bigl\|\bm{x}^{(\bm{s}_j)}_i - \bm{\mu}\bigr\|^2
    \end{aligned}
    \label{eq: original_variance}
\end{equation}
where $N$ is the number of generated slate samples, and each slate $\bm{s}_j$ is sampled from $P_\theta(\bm{s}|\cdot)$.
Note that $\bm{\mu}$ is the average of all $NK$ generated items, and it depends on the input constraint.
If the generative model is personalized, then the user is included in the input of $P_\theta$ and the generation process will first run $N$ times for each user to give personalized variance estimation, then the estimations are averaged for all users.

Let $\bm{\mu}^{(\bm{s})}$ be the average item of slate $\bm{s}$:
\begin{equation}
    \bm{\mu^{(\bm{s})}} = \frac{1}{K}\sum_{i=1}^K \bm{x}^{(\bm{s})}_i\label{eq: slate_average}
\end{equation}
then each slate variance in Eq.\eqref{eq: original_variance} can be decomposed into:
\begin{equation}
    \begin{aligned}
   &  \sum_{i=1}^K\|\bm{x}^{(\bm{s}_j)}_i - \bm{\mu}\|^2 =  \sum_{i=1}^K\|\bm{x}^{(\bm{s}_j)}_i - \bm{\mu}^{(\bm{s}_j)} + \bm{\mu}^{(\bm{s}_j)} - \bm{\mu}\|^2\\
   & = \bigg(\!\sum_{i=1}^K\!(\bm{x}^{(\!\bm{s}_j)\!}_i {-} \bm{\mu}^{(\!\bm{s}_j\!)})\!^\top\!(\bm{x}^{(\!\bm{s}_j\!)}_i {-} \bm{\mu}^{(\!\bm{s}_j\!)}) {+} \!
    \sum_{i=1}^K(\bm{\mu}^{(\!\bm{s}_j)\!} {-} \bm{\mu})\!^\top\!(\bm{\mu}^{(\!\bm{s}_j\!)} {-} \bm{\mu})\\
    &  + 2(\bm{\mu}^{(\bm{s}_j)} - \bm{\mu})^\top\sum_{i=1}^K(\bm{x}^{(\bm{s}_j)}_i - \bm{\mu}^{(\bm{s}_j)})\bigg)\\
\end{aligned}
\end{equation}
Since the last term has $\sum_{i=1}^K(\bm{x}^{(\bm{s}_j)}_i - \bm{\mu}^{(\bm{s}_j)}) = 0$ (from Eq.\eqref{eq: slate_average}), it simplifies the total item variance as:
\begin{equation}
    \mathrm{Cov}(\bm{x}){=} \lim_{\scaleto{N\rightarrow\infty}{4pt} }\! \frac{1}{N}\sum_{j=1}^{N}
    \bigl\|\bm{\mu}^{(\bm{s}_j)} - \bm{\mu}\bigr\|^2 + \frac{1}{NK}\sum_{j=1}^{N}\sum_{i=1}^K\bigl\|\bm{x}^{(\bm{s}_j)}_i - \bm{\mu}^{(\bm{s}_j)}\bigr\|^2
\label{eq: variance_decompose}
\end{equation}
where the first term describes the \textbf{slate-mean variance} and the second term describes the \textbf{intra-slate variance}.
Each of the two terms provides a lower bound for the total item variance, and conversely, the total item variance Eq.\eqref{eq: original_variance} gives an upper bound for either term.
A useful conclusion we can derive from this is that models good at one of the two terms in Eq.\eqref{eq: variance_decompose} may not be the one that achieves the best total item variance.
On one hand, models with good intra-slate variance may still provide repeating slate with the same $\bm{\mu}^{(\bm{s}_j)} = \bm{\mu}$, which results in extremely low slate-mean variance.
On the other hand, models with good coverage of item across slates may still have repeated items (in the most extreme case, $\bm{x}_i^{(\bm{s}_j)}=\bm{\mu}^{(\bm{s}_j)}$ when all items are equal) inside each slate inducing reduced intra-slate variance.
Intuitively, we would like to make both slate-mean variance and intra-slate variance sufficiently large in order to support good total variance. 
Thus, the evaluation protocol should at least include two of the metrics among total item variance, slate-mean variance, and intra-slate variance.

\subsection{The Two-Phase Generation Framework}\label{sec: framework}

\begin{figure}[t]
    \centering
    \includegraphics[width=\linewidth]{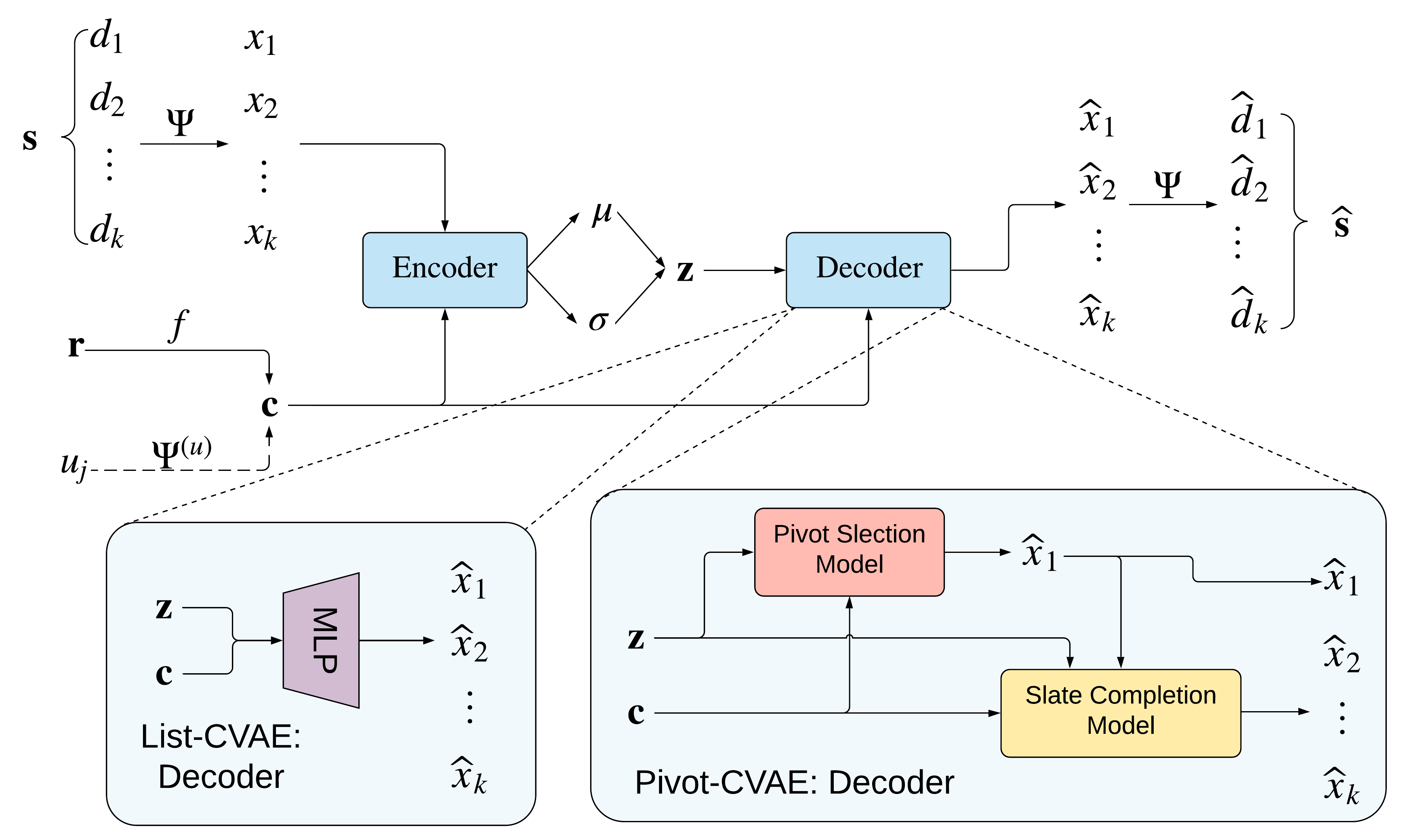}
    \caption{Structure of the generative framework during training. 
    $\bm{s}$ is the input slate of size $k$. 
    $\bm{r}$ is the user response vector of the input slate.
    $\widehat{\bm{s}}$ represents the output slate inferred by decoder.
    $\Psi$ and $\Psi^{(u)}$ extract pretrained embeddings for items and users, respectively.
    }
    \label{fig: model_fig}
\end{figure}

We seek to enforce slate variation when CVAE model provides over-concentrated recommendations (i.e., the large $\beta$ case of RCD).
A straightforward solution is to perturb the generated slate by considering each position as a separate ranking model.
However, this post-generation perturbation is very hard to control and always takes the risk of significant downgrade of recommendation accuracy (detail in Appendix \ref{ap: perturbation}), due to the large perturbation space and the ignorance of the positional bias and item relations.
With this in consideration, we turn to pre-generation perturbation and propose a simple and effective CVAE framework to mitigate the problem.
In general, we separate the original generative process into two steps:
\begin{equation}
\begin{split}\label{eq: decoder}
    P_{\theta}(\bm{s}|\bm{z},\bm{c}) & = P_{\theta}(d_1,\dots,d_K|\bm{z},\bm{c}) \\
    & = P_{\theta}(d_2,\dots,d_K|d_1,\bm{z},\bm{c})P_{\theta}(d_1|\bm{z},\bm{c})
\end{split}
\end{equation}
That is, the framework first uses a \textit{pivot selection model} $P_{\theta}(d_1|\bm{z},\bm{c})$ to select an adequate pivot item for a fixed slate position (here $d_1$ means we always generate the first appearing item in the slate).
Then with this pivot item as additional condition, a \textit{slate completion model} $P_{\theta}(d_2,\dots,d_K|d_1,\bm{z},\bm{c})$ generates the rest of the items for the slate. 
With this separation, we can avoid RCD by enforcing variation of resulting slates through perturbation in the first stage, and use the second phase to clean up the mess if it has made a bad choice of pivot.
As illustrated in Figure \ref{fig: model_fig}, the pivot controller is only applied to the generative decoder.
Compared to the standard VAE model, little has to be nudged in the encoder $Q(\bm{z}|\bm{s},\bm{c})$ since it already has the potential to encode any intra-slate pattern.

\textbf{Picking Pivot Item for the Slate: }\label{sec: pivot_selection_model}
$P_{\theta}(d_1|\bm{z},\bm{c})$ will predict an item as the pivot, based on this, the slate completion model will fill in the rest of the slate according to the pivot.
In other words, the goal of this part is to find the best item for a certain position in the slate, based on the encoding $\bm{z}$ and constraint $\bm{c}$.
It first generates an ``ideal'' latent item embedding $\widehat{\bm{x}}_1$, and then applies dot product with all item embeddings in $\Psi$ to find the closest item as the $d_1$.
The minimization of the reconstruction term can be achieved by optimizing the cross entropy with softmax.
In practice, we also use down sampling~\cite{Mikolov:nips2013:word2vec} to reduce the computational cost and alleviate over-fitting on the training set.
Readers may notice that this part can be treated as a typical ranking model and thus any learning-to-rank framework is suitable for its training, only that one instead of many items are selected at a time.

Similar to sequential modeling, the training of $P_\theta(d_1|\bm{z}, \bm{c})$ is made independent of the later slate completion model, and in both training and inference, this pivot selection phase allows perturbation which improves the item variation.
Yet, perturbation inevitably causes information loss and downgrades the recommendation accuracy.
Theoretically, taking the simplest assumption that item interactions are directional and are all binary relations, then there are at most $K(K-1)$ such interactions between items for a slate of size $K$.
This separation and the introduction of perturbation mean that our model neglects $K-1$ of them (from $K-1$ remaining items towards the pivot).
% For more complicated item relations, it could mean more loss.
Even though, in our experiments, we find that this pre-generation perturbation can improve item variety more significantly with only a minor loss of accuracy compared to post-generation perturbations, which means that the later slate completion model is able to correct the slate according to the perturbed pivot.
Additionally, we suggest to pick one pivot instead of more in this phase, since for any $1<k^\prime<K$ (in the binary relation case), when choosing $k^\prime$ pivots, the number of missing relations will be $(K-k^\prime)k^\prime \geq K-1$, which indicates more loss of information and recommendation accuracy.

\textbf{Slate Completion with a Given Pivot Item: }\label{sec: slate_completion_model}
After the selection of the pivot, the goal of the slate completion model 
\begin{equation}
    P_{\theta}(d_2,\dots,d_K|d_1,\bm{z},\bm{c})
    \label{eq: slate_model}
\end{equation} 
is to learn to fill up the remaining items that can satisfy the desired constraint $\bm{c}$.
A forward pass will take as input the selected pivot $\widehat{d}_1$, the encoding $\bm{z}$ (which is the output of $Q$ if training, output of the conditional prior $P_{\theta}(\bm{z}|\bm{c}^*)$ if inference, as in VAE Eq.\eqref{eq: cvae_loss}), and the constraint $\bm{c}$, then output a set of ``best'' latent item embeddings $\widehat{\bm{x}}_2, \dots, \widehat{\bm{x}}_K$ for each of the remaining slots in the slate.
After generating these latent embeddings, it will find for each of the $\widehat{\bm{x}}_i$ the nearest neighbor in the candidate set $\mathcal{D}$ through dot product similarity.
Similar to that in the pivot selection model, we can again apply cross-entropy loss with softmax and negative sampling during training.
Note that this is the final generation stage and it does not employ perturbation.

Compared to inference time when the model can only use the inferred $\widehat{d_1} \sim p_\theta(d_1|\bm{z},\bm{c})$ from the pivot selection model, during training, there is another valid choice of the pivot - the ground truth item in the data.
We find that the later choice achieves the same performance but usually exhibits faster convergence.
Thus, we adopts the ground truth item $d_1$ for the input of the slate completion model during training in our experiments, and if perturbation, we calculate item similarities based on the ground truth instead of the inferred item embedding.
Additionally, when the pivot is perturbed during training, the slate completion model tends to learn a ``denoised'' intra-slate patterns which may results in a slate that is more accurate but with less variation, compared to training without perturbation, as we will discuss in section \ref{sec: improve_variance}.

\section{Experimental Setting}\label{sec: experiments}

\subsection{Real-world Datasets}\label{sec: datasets}

We conducted experiments
\footnote{Code link: https://github.com/CharlieFaceButt/PivotCVAE} on two real-world datasets.
The first is \textbf{YOOCHOOSE}
\footnote{https://2015.recsyschallenge.com/challenge.html} from RecSys 2015 Challenge 
and we follow the same reprocessing procedure as~\cite{jiang2018beyond}.
The resulting dataset contains around 274K user slate-response pairs.
Note that there is no user identifier involved in this dataset, so our second dataset is constructed from the MovieLens 100K\footnote{https://grouplens.org/datasets/movielens/100k/} dataset.
We split user rating sessions into slates of size 5 and consider the rating of 4-5 as positive feedback (with label 1) and 1-3 as negative feedback (with label 0).
The resulting distribution of slate responses (Figure \ref{fig: slate_resp_dist} in Appendix~\ref{sec:slate_res}) is similar to that in the Yoochoose dataset.
We consider two versions of this dataset: \textbf{ML (User)} and \textbf{ML (No User)} to investigate how the presence of user affects the generative results.
Compared to ML(User), the ML(No User) dataset ignores user IDs like Yoochoose data.
Since both datasets are skewed towards slates with 0 and 3 clicks, we augment the records of 1,2,4, and 5 clicks by random repetition until each group has at least half the size of the largest response type.
Note that these offline log data have limited feasibility for evaluation since they cannot provide accurate estimations for unseen records.
Thus, an additional user response model $R: \mathcal{D}^K \rightarrow \{0,1\}^K$ is trained (with binary cross-entropy loss) to fulfill the role of ``ground truth'' user feedback.

\subsection{Simulation Environment Settings}

To observe how generative models behave for unseen slates under different environment settings and to investigate the difference between slate generation metrics and traditional ranking metrics, similar to existing works~\cite{jiang2018beyond, ie2019slateq}, we employ simulations with plugins of positional biases and item interactions.

The primary goal of the simulated environment is to model $R(\bm{r}|\bm{s},\bm{u})$ that predicts the users' true responses given slate $\bm{s}$.
And for each of the simulators described in this section, the final response for each item $d_k$ is sampled by Bernoulli distribution with click probability $\mathcal{I}(d_k, j)$, which represents user $j$'s interest for $d_k$:
\begin{equation}
    r_k = R(r_{kj}|d_k,j) \sim \mathrm{Bernoulli}(\mathcal{I}(d_k, j))
\end{equation}
Thus, the click behavior follows Poisson binomial distribution, and the expectation of the number of clicks is:
\begin{equation}\label{eq: sim_sampling}
    \mathbb{E}\Bigl[\sum_{k=1}^K r_k\Bigr] = \sum_{d_k \in \bm{s}} \mathcal{I}(d_k, j)
\end{equation}
We tune the resulting distribution with proper setting (details in appendix \ref{ap: simulation}) so that it coincides with that of real-world datasets.

Specifically, each simulation is built based on a basic
\textbf{User Response Model (URM)}, which only considers point-wise user-item responses like the matrix factorization model.
By adding awareness of positional bias and multi-item relations, we obtain \textbf{URM\_P} (P stands for positional bias) and \textbf{URM\_P\_MR} (MR stands for multi-item relations), respectively. 
The URM\_P\_MR consists of a coefficient $\rho$ for the weight of the multi-item relations.
As a special case, setting $\rho=0$ will tell the simulation to include no item relations and the environment will reduce to \textbf{URM\_P}.
The details of each simulation environment are given in Appendix \ref{ap: simulation}.

\textbf{Simulation Data: } We set up three URM\_P\_MR environments ($|\mathcal{D}| = 3,000$, $|\mathcal{U}| = 1,000$) with different values of $\rho \in \{0, 0.5, 5.0\}$.
Note that there is no need to train a response model from the generated dataset like that for real-world datasets.
Conversely, we generate a training set of 100,000 slates from each environment.
The number of slates for all types of user responses are also balanced similar to that of real-world datasets.
The user and item embeddings are assumed explicit and free to use in the training of the recommendation model.
Here we expect readers to distinguish these simulations from those used in Reinforcement Learning (RL)-based recommendation models, because the generative model does not interact with the simulated environment for rewards during training.
In other words, the generative model is training offline and the simulators are only used for evaluation purposes.

\subsection{Model Specification}\label{sec: baselines}

We denote our two-step generative process as Pivot-CVAE.
For Pivot-CVAE model, perturbation of $d_1$ can be applied either on training phase or inference phase, inducing 4 possible variations:
where ``GT'' represents that the model uses Ground Truth item during training, ``PI'' represents that the model uses Pivot Item during inference, and ``S'' means the item applies perturbation.
For all perturbation, we adopt sigmoid dot-product between item embeddings as similarity and sample according to multinomial distribution so that it can capture user interests.

\begin{table}
    \centering
    \caption{Pivot-CVAE variations}
    \begin{tabular}{ccc}
        \toprule
        \multirow{2}{*}{Models} & \multicolumn{2}{c}{perturbation of $d_1$}\\
        \cmidrule(lr){2-3}
         & training time & inference time \\
        \midrule
        Pivot-CVAE (GT-PI) &  & \\
        Pivot-CVAE (SGT-PI) & \checkmark & \\
        Pivot-CVAE (GT-SPI) &  & \checkmark\\
        Pivot-CVAE (SGT-SPI) & \checkmark & \checkmark\\
        \bottomrule
    \end{tabular}
    \label{tab: pivot_cvae_variations}
\end{table}

\textbf{Baselines:}
We include the \textbf{List-CVAE} model \cite{jiang2018beyond} as an example of VAE and build its non-greedy version (denote as \textbf{Non-Greedy List-CVAE}) that conducts post-generation perturbation.
That is, after the generation of the slate, the item $d_1$ (in the same position as the pivot of Pivot-CVAE) is perturbed by sampling from $\mathcal{D}$.
Again, we apply sampling based on multinomial distribution of sigmoid dot product similarity.
We also include
biased \textbf{MF} \cite{koren2009matrix}
and \textbf{NeuMF} \cite{he2017neural} as representatives of discriminative ranking models.
% We apply the pointwise learning scheme using binary cross-entropy loss on each click/no-click label. 
% One can also use pairwise learning with negative sampling, but the resulting model will always be pointwise for each user-item pair.
% While MF can learn the basic interactions between users and items, SVD++ can explicitly take into account the user history.
% NeuMF on the other hand is included as a neural network baseline that may be able to find complex user-item behaviors.
In order to engage generative recommendations that can explore items other than the top items, we extend these discriminative methods into \textbf{Non-greedy MF/NeuMF} by applying the same perturbation method on $d_1$ as that in Non-greedy List-CVAE and Pivot-CVAE.
To compare the item variance with intra-slate variance, we include the widely adopted \textbf{MF-MMR} \cite{carbonell1998mmr} as a representative diversity-aware method. 
It re-ranks the items proposed by the pre-trained biased MF model based on the following modified score:
\begin{equation*}
    \mathrm{score}(d) = \lambda\, \mathrm{sim}(d,j) + (1-\lambda)\max_{d_i\in\bm{s}}\mathrm{sim}(d_i,d)
\end{equation*}
where slate $\bm{s}$ starts from an empty set and choose the item with the best MMR score in each step until the slate size is $K$.
$\mathrm{sim}(d,j)$ represents the item's original ranking score given by the base MF model, and $\mathrm{sim}(d_i,d)$ is the item's similarity to the $i$-th item that has already been added to the list $\bm{s}$.

In our experiment, we adopt two-layered network with 256 dimensional hidden size for each encoder, decoder, $P_\theta(d_1|\bm{z},\bm{c})$, the slate completion model $P_\theta(d_2,\dots,d_K|d_1,\bm{z},\bm{c})$, and the MLP component in NeuMF.
In terms of the performance of CVAE-based models, we found that it is relatively insignificant to change the width or depth of the encoder and decoder network as long as they are large enough.
The user and item embedding size for all datasets and simulations are fixed to 8, and the size of $\bm{z}$ is set to $m=16$.
The slate size is $K = 5$, which means the size of the condition input $\bm{c}$ of CVAE-based model is $K+1=6$ (without user condition) as described in the first paragraph of section \ref{sec: slate_recommendation}.
All models are optimized by Adam with stochastic mini-batches (batch size of 64), and we use grid search to find the best learning rate (0.0001 for List-CVAE and Pivot-CVAE, 0.0003 for MF and NeuMF) and weight decay (0.0001 for all models).
For MF and NeuMF, we follow the standard LTR paradigm with point-wise binary cross-entropy loss and assign 2 random negative samples of each record to optimize these models until their ranking performance converges in the validation set.
For MF-MMR, we use sigmoid dot-product as item similarity and set $\lambda=0.5$.
During training of generative models, the softmax function on each slot in a slate is associated with 1000 negative samples for Yoochoose, and 100 negative samples for MovieLens and simulation environments.

\subsection{Evaluation Protocol}\label{sec: evaluation}

For all datasets, we randomly split them into train, validation, and test sets following the 80-10-10 holdout rule.
And we run each experiment five times to obtain the average performances.
We consider two major evaluation metrics based on interactive environment $R(\bm{r}|\bm{s})$: slate accuracy and slate variation.
And for the illustration of why ranking metrics on test set is invalid for evaluation of generative models, we further include discriminative ranking metrics.

\textbf{Slate Accuracy Metric:}
The primary metric, following the evaluation setting of \cite{jiang2018beyond}, is the \textit{Expected Number of Clicks} (ENC) which is calculated as:
\begin{equation*}
    \mathbb{E}\left[\sum_{k=1}^{K}r_k\right] = \sum_{\bm{s} \in \mathcal{D}^K}P(\bm{s})\mathbb{E}\left[\sum_{k=1}^K r_k|\bm{s}\right]
\end{equation*}
where $\bm{r}_k |\bm{s}$ is a random variable modeled by $R(\bm{r}|\bm{s})$, and $P(\bm{s})$ is the probability of generating $\bm{s}$.
Similar to the variation evaluation described in section \ref{sec: variance}, we can approximated this metric by sampling techniques.
This metric is exactly the ultimate goal of the optimization and does not involve any test set compared to traditional ranking metrics.
For simulation, combining Eq.~\eqref{eq: sim_sampling}, it becomes:
\begin{equation*}
    \mathbb{E}\left[\sum_{k=1}^{K}r_k\right] = \sum_{\bm{s} \in \mathcal{D}^K} P(\bm{s}) \sum_{d_k \in \bm{s}} \mathcal{I}(d_k, j)
\end{equation*}
And for real-world dataset, we train $R(\bm{r}|\bm{s})$ ($R(\bm{r}|\bm{s},u)$ if user IDs are presented) with point-wise binary cross entropy minimization.

\textbf{Slate Variation:} 
This metric reveals the severance of the ``over concentration'' in RCD and the generation pitfall of limited slate prototypes. %, and exploration magnitude.
As described in section \ref{sec: variance}, we use total item variance and intra-slate variance metrics in our evaluation.
Notably, the variance of $\bm{z}$ directly models the slate variance, but it is unique in VAE-based generative models.
In order to form comparison with non-VAE models, we use item \textit{Coverage} \cite{good1999combining} as the item variance metric and \textit{Intra-List Diversity} (ILD)~\cite{ziegler2005improving,zhou2010solving} as an approximation of the intra-slate variance.
Item coverage estimates the proportion of unique items in $\mathcal{D}$ that can appear after several times of generations.
Obviously, LTR models are deterministic so will always cover only $5/|\mathcal{D}|$ of the items without perturbation.
Intra-list diversity is based on Intra-List Similarity (ILS) \cite{ziegler2005improving}:
\begin{equation*}
    \mathrm{ILD} = 1 - \mathrm{ILS}(\bm{s}) = 1 - \sum_{d_i \in \bm{s}}\sum_{\substack{d_l \in \bm{s}\\d_l \neq d_i}} g(\bm{v}_i^{\top} \bm{v}_l)
\end{equation*}
where the similarity measure $g$ between $d_i$ and $d_l$ in the slate is based on the dot product of their item embeddings.

\textbf{Ranking Metrics: } 
We agree with \cite{jiang2018beyond} that it is inadequate to use traditional offline ranking metrics to evaluate generative models, as we will discuss in section \ref{sec: gen_vs_dis}, these metrics behave differently on a test set compared to that on a interactive user response environment.
Even though, it is still reasonable to compare these metrics among generative models.
Specifically, we calculate slate \textit{Hit Rate} and slate \textit{Recall} considering each slate as a ranking list.
It is considered as a ``hit'' if an item in the ground truth slate with positive feedback is recommended.
And the slate recall considers each slate as a user history instead of the combined user history across slates.
Note that in Yoochoose and ML, user identifiers are absent, so we assume a universal user for all slates during training.

In summary, we conduct two types of evaluation: 1) recommendation performance (slate accuracy and variance metric) on $R(r|s)$ as main evaluation, and 2) ranking metric on the test set.
Due to the stochastic nature of generative models (List-CVAE, Pivot-CVAE, and all Non-greedy models), the evaluation of each metric is calculated based on $N$ sampled outputs (correspond to section \ref{sec: variance}).
Note that $N$ cannot be too small or else it will not provide accurate and stable estimation of the true value.
In the meantime, it can neither be too large, otherwise the model would exhibit indistinguishably high item coverage (i.e. it may simply generate all items in $\mathcal{D}$ given sufficient number of samples).

%%%%%%%%%%%%%%%%%%%%%%%%%%%%%%%%%%%%%%%%%%%%%%%%
%                  Results                     %
%%%%%%%%%%%%%%%%%%%%%%%%%%%%%%%%%%%%%%%%%%%%%%%%

\section{Results and Discussions}\label{sec: results}

\begin{figure*}[t]
    \centering
    \includegraphics[width=0.92\textwidth]{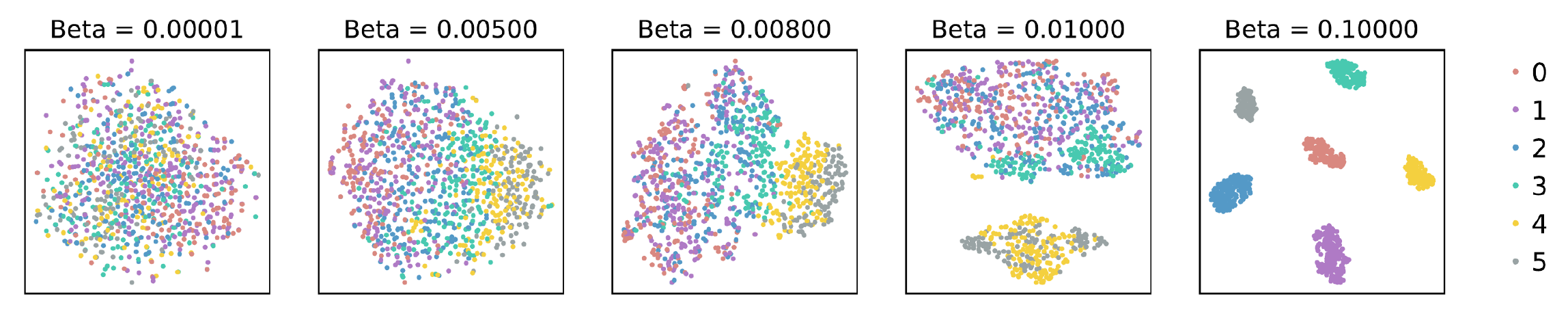}
    \caption{The slate encoding TSNE plots of List-CVAE on Yoochoose.
    The first plot correspond to over-reconstruction case, the last corresponds to over concentration case, and the middle plots correspond to the ``elbow'' case.}
    \label{fig: yoochoose_beta_tsne}
\end{figure*}

\begin{figure}[t]
    \centering
    \includegraphics[width=0.46\textwidth]{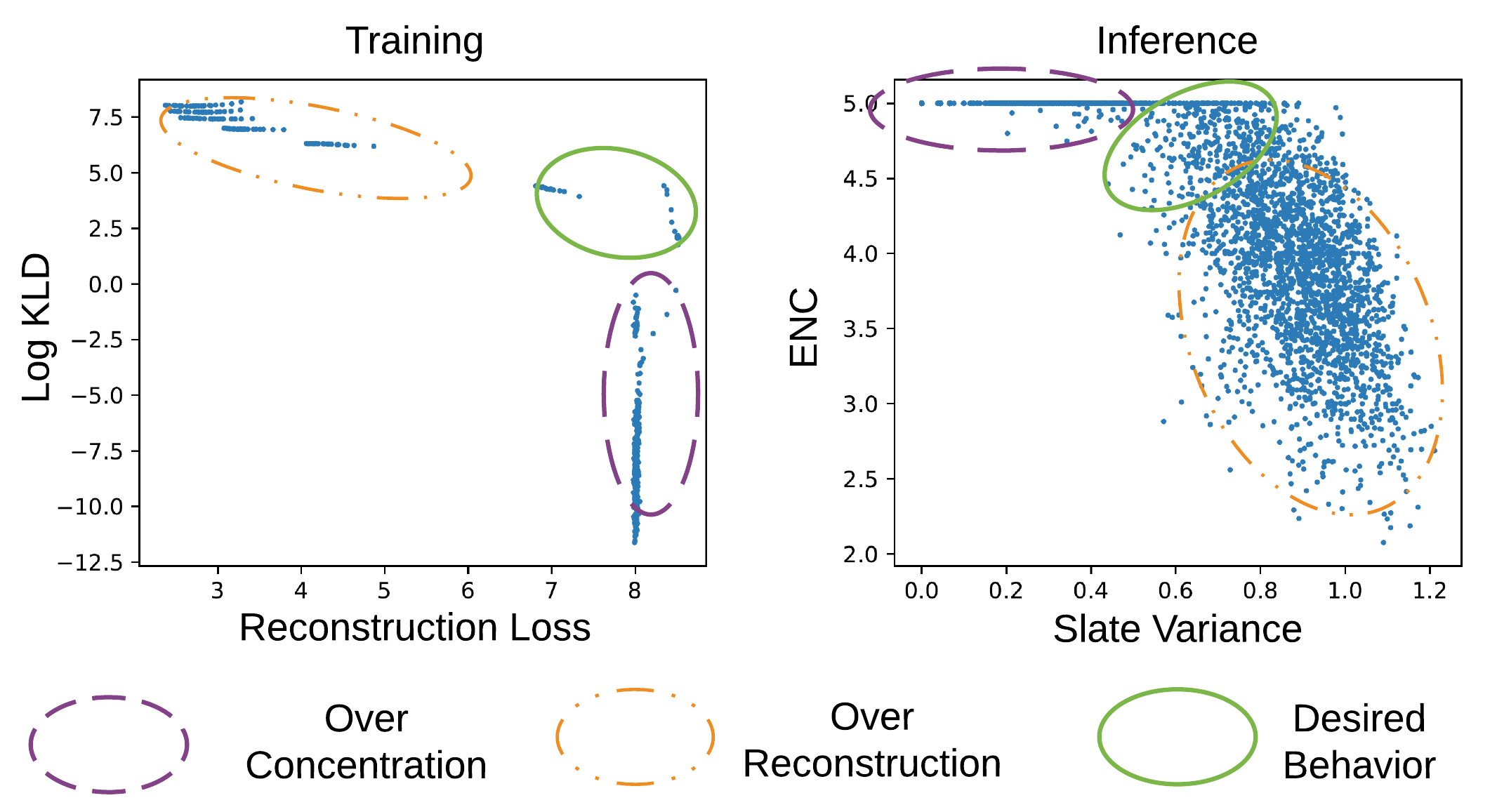}
    \caption{Training loss behavior (left) and recommendation performance (right) of RCD on the Yoochoose Data.
    Each point in the left panel represents the average result of slates in one training epoch of a model.
    Each point in the right panel represents a certain generated slate.
    Here, we use ILD as slate variation, ENC as accuracy metric. 
    }
    \label{fig: cvae_tradeoff}
\end{figure}

\subsection{\mbox{\hspace{-5pt}The Reconstruction-Concentration Dilemma}}\label{sec: rcd}

We consider the search space of $\beta\in [0.00001,30.0]$ (chosen uniformly on $\log\beta$ space) and for each setting we train List-CVAE and all Pivot-CVAE models until convergence of ENC on $R(\bm{r}|\bm{s})$.
When evaluation, we generate $N=500$ slates from each trained model and calculate the average as described in section \ref{sec: evaluation}.
In Figure \ref{fig: cvae_tradeoff}, we plot the RCD pattern of List-CVAE on Yoochoose dataset, and we have observed the same pattern in MovieLens 100K and all simulation environments.

In cases where $\beta$ is small, CVAE becomes biased towards learning the reconstruction term of Eq.~\eqref{eq: cvae_loss} as illustrated by the yellow dot-dashed circle in the left subplot of Figure \ref{fig: cvae_tradeoff}.
And because of the subdued regularization from the KL term, the encoding distribution of $\bm{z}$ becomes less aligned with that of the predefined prior.
When setting the prior $P_\theta(\bm{z}|\bm{c})$ as isotropic standard Gaussian, we observe that the means of the inferred $\bm{z}$ are often significantly deviated from $\bm{0}$ and the variances $\textrm{var}(\bm{z})$ are far from $\bm{1}$.
Though it successfully learns and generates the slates in the dataset during training, there is no guarantee on the effectiveness of the sampled $\bm{z}$ during inference.
In other words, the distribution of generated slates is close to a random selection on the observed dataset.
As shown in the yellow dot-dashed circle in the right subplot of Figure \ref{fig: cvae_tradeoff}, we observe low ENC and high variance during inference.
% entire candidate set $\mathcal{D}$.

On the contrary, in the over-concentration case where $\beta$ is rather large, the KL term plays a more important role in the learning.
The slate encoding $\bm{z}$ indeed is more aligned with the prior, ensuring the sampling effectiveness, and consequently able to generate satisfactory slates during inference.
Yet, it is less capable of encoding information that is necessary to reconstruct the slates.
When the model learns that $\bm{z}$ is reluctant to encode corresponding slates, the generator tends to ignore $\bm{z}$ and focuses on the condition $\bm{c}$.
Since $\bm{c}$ alone does not contain any variational information about slates, the model will only be able to output several biased ``slate prototypes'' (as illustrated in Appendix \ref{ap: rcd}, second row of Figure \ref{fig: cvae_beta_tunning}).
An alternative analysis of the slate encoding $\bm{z}$ of List-CVAE is given in Figure \ref{fig: yoochoose_beta_tsne}.
It shows that with large $\beta$, slate encoding becomes disjoint according to the ground truth number of clicks, which means that slate encoding tends to gather around its corresponding prototype given by the prior.
This is undesirable since the model cannot infer slates outside the cluster, which results in the lack of variety in recommendations.

Besides, we notice that in the training data a lot of repeated clicks appear in the click and/or purchase sessions in Yoochoose data. This makes the RCD problem even worse since the same item is repeatedly recommended even within the same slate, inducing low intra-slate variance.
We observed that RCD exists even with $\beta$-annealing~\cite{bowman2015generating}, disabled condition (reduce CVAE to VAE), and constrained variation (only fix the variation of $z$, but not the mean).
These observations indicates that RCD problem may exist for a broad range of generative models.

\subsection{The Narrow ``Elbow'' of CVAE}

\begin{figure}[t]
    \centering
    \includegraphics[width=0.48\textwidth]{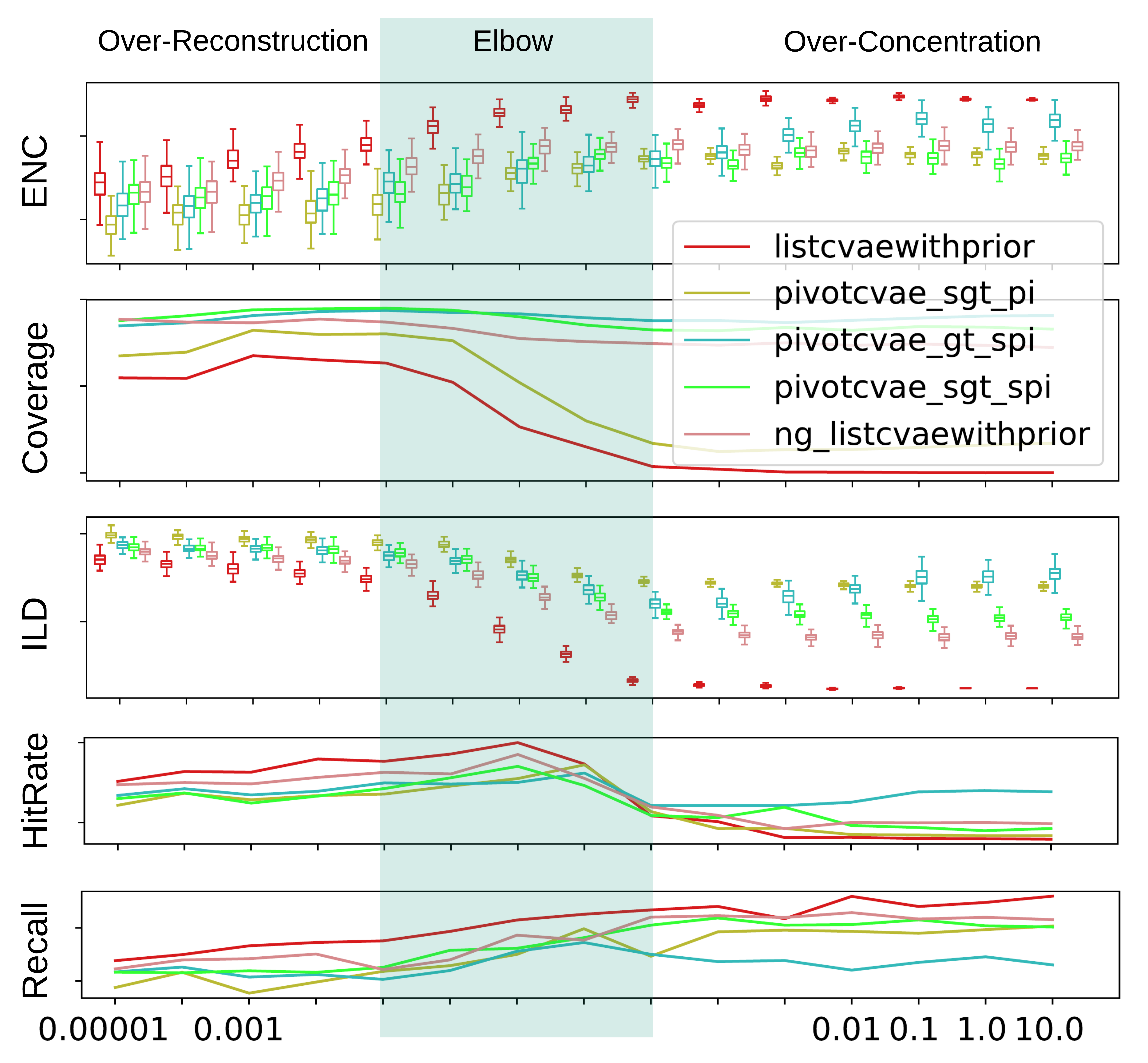}
    \caption{Performance on ML (No User) data.
    ``listcvaewithprior'' represents the List-CVAE and
    ``ng\_listcvaewithprior'' corresponds to the Non-Greedy List-CVAE. 
    % {\color{blue} can you add spaces to the ``listcvaewithprior'' in the figure legend?}
    }
    \label{fig: all_metric_movielens}
\end{figure}

Though neither of the extremes appears to be a good choice for recommendation, we find that there exists a very narrow region 
% (e.g. the green elbow circle in Figure \ref{fig: cvae_tradeoff})
in between, where models can provide feasible outputs.
In Figure \ref{fig: all_metric_movielens}, we show case the results of all metrics on ML(No User) data for generative models across different $\beta\in[0.00001,10.0]$.
X-axis represents the setting of $\beta$ and note that results for different $\beta$s correspond to different models that are separately trained and evaluated.
For ENC and ILD metrics, we use box plot to better demonstrate the distribution of generated slates.

\begin{figure*}[]
    \centering
    \includegraphics[width=0.9\textwidth]{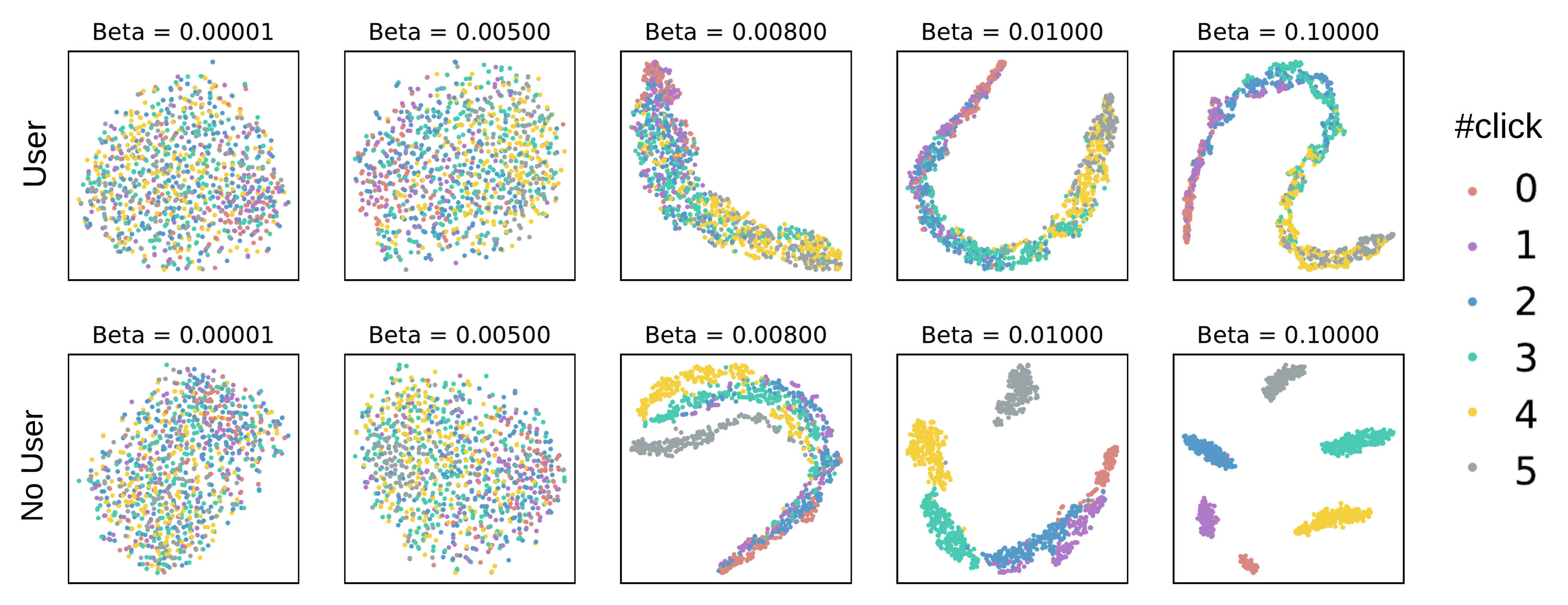}
    \caption{The slate encoding TSNE plots of List-CVAE on MovieLens datasets.
    When user identifier is presented, the encoding forms more fine-grained clusters that is no longer disjoint between one another.}
    \label{fig: movielens_tsne}
\end{figure*}

We summarize three trends of model behavior when increasing the value of $\beta$ as follows: 
\begin{itemize}
    \item For model training, the converged reconstruction loss gets worse while the KLD loss gets better; 
    \item When inference, the accuracy measure ENC starts to boost but the variation metric of the generated slates drops;
    \item $\bm{z}$ starts to show clustering behavior under the regulating prior and the clusters will become crisper along with the transition as shown in Figure \ref{fig: movielens_tsne}.
\end{itemize}

This transitional behavior indicates that models in this intermediate region can to some extent cover the variety of slates in the data while provide moderate accuracy performance.
To better show the detailed transitional behavior of the feasible region, we include a more fine-grained search space for $\beta\in[0.001,0.01]$ and highlight it with shaded green in Figure \ref{fig: all_metric_movielens}.
However, in the experiment of both real-world datasets and all simulation datasets, we found that this transition happens within a very small region (at most $30\%$ of the $\log \beta$ search space or equivalently $2\%$ of the $\beta$ search space), while the search space in our experiment is $\beta \in \left[0.00001, 30.0\right]$

Additionally, we observe that this transitional region consistently gives good test set ranking performance (both hit rate and recall) compared to other choices of $\beta$.
The two extreme cases outside the ``elbow'' region do not always reveal a decreasing hit rate and an increasing recall on the test set as in Figure \ref{fig: all_metric_movielens}, but the best ranking performance usually appears in one of the two sides.
Intuitively, the generative model should be able to maximize the likelihood of the test set in addition to the training set.
Following the same derivation of Eq.\eqref{eq: cvae_loss}, this would require both the ability to reconstruct the slate information and the ability to satisfy the constraint.
This can only be observed in this transitional region if the slate variation is not enforced, since the two extremes only possess one of the two characteristics.
Note that this ranking performance can only serve as indicators to compare generative models, and it is incomparable between deterministic ranking models and stochastic generative models.
As we will discuss in section \ref{sec: gen_vs_dis}, the stochastic generation process explores and proposes various good slates in the view of the user $R(\bm{s}|\bm{c})$, and may not necessarily pin-point the data in the test set thus it is typically not favored by this kind of metrics.

\subsection{Controlling Slate Variation}\label{sec: improve_variance}

\begin{table}[t]
\renewcommand{\arraystretch}{0.97}
    \centering
    \caption{Model Performance on User Feedback $R(r|s)$ of datasets with user IDs. All results are significant ($p<0.05$) and the overall best are the bold scores while the best among generative models are underlined.}
    \begin{adjustbox}{max width=\linewidth}
    \begin{tabular}{ccccc}
    \toprule
         & ML(User) & URM\_P & $\substack{\textrm{URM\_P\_MR}
         \\(\rho=0.5)}$ & $\substack{\textrm{URM\_P\_MR}\\(\rho=5.0)}$\\
        \bottomrule
        \toprule
        \multicolumn{5}{c}{A: Expected Number of Click (ENC)}\\
        \midrule
        % \midrule
        MF & 3.246 & \textbf{3.353} & 3.870 & 4.961 \\
        NeuMF & 3.197 & 3.344 & 3.810 & 4.938 \\
        MF-MMR & 2.400 & 3.243 & 3.725 & 4.617 \\
        \midrule
        Non-Greedy MF & 2.950 & 3.315 & 3.755 & 4.869 \\
        Non-Greedy NeuMF & 3.020 & 3.303 & 3.730 & 4.819 \\
        \midrule
        List-CVAE & \underline{\textbf{3.579}} & 3.237 & 3.924 & \underline{\textbf{4.971}}\\
        \midrule
        Non-Greedy List-CVAE & 3.285 & 3.262 & 3.883 & 4.777\\
        Pivot-CVAE (SGT-PI) & 3.376 & \underline{3.274} & \underline{\textbf{3.934}} & 4.944 \\
        Pivot-CVAE (GT-SPI) & 3.252 & 3.226 & 3.711 & 4.622\\
        Pivot-CVAE (SGT-SPI) & 3.152 & 3.270 & 3.816 & 4.704\\
        \bottomrule
        \toprule
        \multicolumn{5}{c}{B: Item Coverage}\\
        \midrule
        MF & 0.003 & 0.002 & 0.002 & 0.002 \\
        NeuMF & 0.003 & 0.002 & 0.002 & 0.002 \\
        MF-MMR & 0.003 & 0.002 & 0.002 & 0.002 \\
        \midrule
        Non-Greedy MF & 0.142 & 0.082 & 0.082 & 0.080 \\
        Non-Greedy NeuMF & 0.141 & 0.082 & 0.081 & 0.080 \\
        \midrule
        List-CVAE & 0.004 & 0.030 & 0.011 & 0.005\\
        \midrule
        Non-Greedy List-CVAE & 0.139 & 0.106 & 0.088 & 0.084\\
        Pivot-CVAE (SGT-PI) & 0.071 & 0.065 & 0.014 & 0.005\\
        Pivot-CVAE (GT-SPI) & \underline{\textbf{0.250}} & \underline{\textbf{0.235}} & \underline{\textbf{0.180}} & \underline{\textbf{0.227}}\\
        Pivot-CVAE (SGT-SPI) & 0.144 & 0.097 & 0.090 & 0.083\\
        \bottomrule
        \toprule
        \multicolumn{5}{c}{C: Intra-List Diversity (ILD)}\\
        \midrule
        MF & 0.206 & 0.031 & 0.035 & 0.036\\
        NeuMF & 0.694 & 0.300 & 0.534 & 0.779 \\
        MF-MMR & 0.287 & 0.230 & 0.193 & 0.227 \\
        \midrule
        Non-Greedy MF & 0.545 & 0.515 & 0.231 & 0.126 \\
        Non-Greedy NeuMF & \textbf{0.836} & 0.576 & 0.644 & \textbf{0.827} \\
        \midrule
        List-CVAE & 0.178 & 0.836 & 0.407 & 0.524\\
        \midrule
        Non-Greedy List-CVAE & 0.428 & 0.864 & 0.572 & 0.664\\
        Pivot-CVAE (SGT-PI) & 0.486 & 0.869 & 0.451 & 0.632 \\
        Pivot-CVAE (GT-SPI) & \underline{0.725} & \underline{\textbf{0.945}} & \underline{\textbf{0.740}} & \underline{0.814}\\
        Pivot-CVAE (SGT-SPI) & 0.551 & 0.856 & 0.600 & 0.637\\
        \bottomrule
    \end{tabular}
    \end{adjustbox}
    \label{tab: full}
\end{table}

We present the results of ENC and variance in Table \ref{tab: full}. Generative models with $\beta=1.0$ are chosen as representatives of the large-$\beta$ case, since we want to observe the improvement of slate variance when models are over-concentrated.
Generative models with small $\beta$ (described in section \ref{sec: rcd}) and post-perturbation methods that change more than one item cannot provide satisfactory user response, so they are not included in the comparison.
We only present results of datasets with user IDs (ML (User) and all simulation environments) so that collaborative filtering models like MF and NeuMF can be compared.
The result of each stochastic model (Non-Greedy models, List-CVAE and Pivot-CVAE models) is calculated by the average of all users' evaluation.
Note that when calculating item coverage and diversity, we consider user-wise instead of the system-wise metric for these datasets.

The List-CVAE baseline achieves the best ENC on ML(User) and URM\_P\_MR environments because it is over-concentrated on the optimal slate prototype, and CF models achieves the best ENC on URM\_P because of the pointwise environment.
All models with item perturbation (Non-greedy List-CVAE, Pivot-CVAE (SGT\_PI), Pivot-CVAE (GT\_SPI), and Pivot-CVAE (SGT\_SPI)) exhibit degraded ENC compared with the original List-CVAE, but significantly improves slate variation (Item Coverage and ILD).
Among models using perturbation, the Pivot-CVAE (GT-SPI) model always achieves satisfactory accuracy with the best slate variety.
We observe this outstanding performance across all datasets, meaning that sampling the pivot during inference (SPI) will induce more variance and explore more choices of item combinations than sampling during training (SGT).
Pivot-CVAE (SGT\_PI) applies perturbation during training but not inference, this allows the model to give more accurate generation with better ENC, but the improvement of item coverage and ILD becomes limited.
Note that it can achieve a similar ILD with Non-Greedy List-CVAE even if there is a huge gap between their item coverages, indicating that SGT\_PI seeks to find good slates with sufficient intra-slate variance but tends to be concentrated slate-wise in exchange for good accuracy.
When applying perturbation on both training and inference as Pivot-CVAE (SGT-SPI), it has similar performance to Non-Greedy List-CVAE.

As shown in Table \ref{tab: full}, generative methods consistently outperform MF and NeuMF on variance metrics, and achieves better ENC on all datasets except for URM\_P where the environment is pointwise.
This indicates that the user responses of real-world datasets like ML(User) are closer to URM\_P\_MR, which contain intra-slate features such as item relations, rather than URM\_P.
Additionally, Non-Greedy MF/NeuMF can improve the item coverage of these LTR models to the level of Non-Greedy List-CVAE baseline (still worse than Pivot-CVAE (GT-SPI)) and Non-Greedy NeuMF even occasionally achieves better ILD performance than Pivot-CVAE (GT\_SPI).
However, they achieve this with greater sacrifice on the ENC.
On the other hand, MF-MMR is able to increase ILD, but its performance is worse than generative models on all metrics.
Moreover, it also shows that a model that improves intra-slate variance does not necessarily improve the total item variance.

\subsection{Personalization Improves Variance}\label{sec: user_vs_nouser}

Different from Yoochoose and MovieLens (No User) Data, the MovieLens (User) and our simulation environments include user ID in the constraints in addition to the ideal response, allowing the model to learn personalized preference of slates.
We plot the distribution of $\bm{z}$ (of List-CVAE) in Figure \ref{fig: movielens_tsne} to show their difference in over-concentration case.
For generative models trained with large $\beta$, instead of having disjoint slate encoding clusters for each type of user response, the presence of user ID in the constraint will guide the model to learn a set of more fine-grained clusters, each of which corresponds to a user.
Note that the same user may have different types of user responses, and a typical user that usually gives a certain type of response also has a higher chance of giving responses of similar types (e.g., a user frequently clicks everything may also frequently click $K-1$ items).
Consequently, user clusters become closer if they give similar types of response and closer response types become partially mixed with each other because of the common users, thus forming a topologically sorted chain in the space, as shown in the right most panels in the first row of Figure \ref{fig: movielens_tsne}.
This property will contribute to the total item variation of the overall system across users, but in a personalized view, the concentration of slate still exists.
As given in Table \ref{tab: full}-A, the user-wise item coverage of List-CVAE is close to that of discriminative ranking models.

%%%%%%%%%%%%%%%%%%%%%%%%%%%%%%%%%%%%%%%%%%%%%%%%
%                 Conclusion                   %
%%%%%%%%%%%%%%%%%%%%%%%%%%%%%%%%%%%%%%%%%%%%%%%%

\section{Conclusion and Expectations}

In this paper, we show that generative models for slate recommendation tasks may fall into the Reconstruction-Concentration Dilemma (RCD), where only a narrow middle region can produce effective recommendations.
We point out that personalization or applying perturbation can enforce variation on the over-concentration case of the dilemma but have limitations.
By separating a pivot selection phase from the generation process, we propose Pivot-CVAE mode that offers better control of the slate variation by perturbation before the generation. 
Our pivot-based approach and the variation evaluation framework can be extended to a wider scope of stochastic generation models such as Generative-Adversarial Networks (GAN) \cite{wang2017irgan}, which we will explore in the future.
Besides, we also find it useful to construct a flexible and comprehensive user-response simulation framework, not only for the purpose of recovering a realistic recommendation environment but also for the need of generating unseen user-item interactions for model training and evaluation, which is essential for generative models as well as causal ~\cite{holland1986statistics, bonner2018causal, mcinerney2020counterfactual} and RL-based models. 
We will extend our framework for training and evaluating these models in the future.


\begin{acks}
We thank Ji Zhang, Qi Dong, and all the reviewers for the constructive discussion.
\end{acks}

\balance

\newpage

\bibliographystyle{ACM-Reference-Format}
\bibliography{main}

\balance

\newpage

\appendix

\section*{Appendix}

\begin{figure*}[t]
    \centering
    \includegraphics[width=0.9\textwidth]{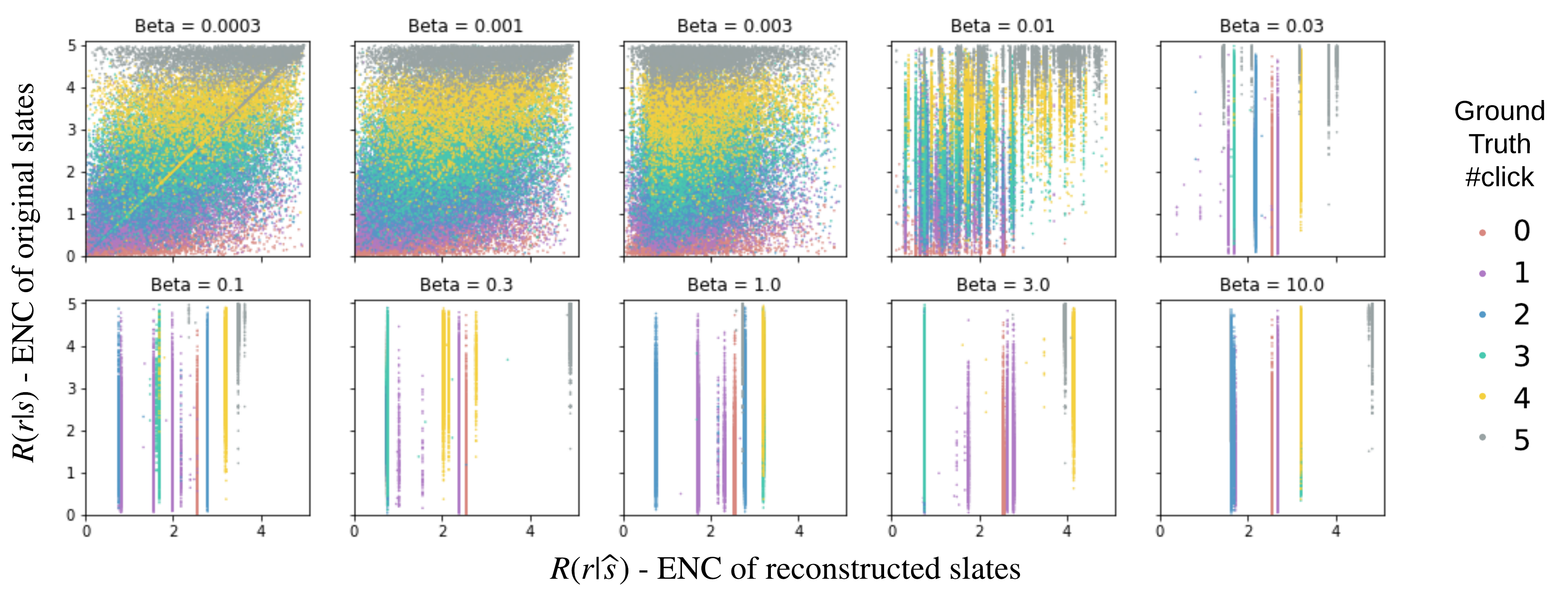}
    \caption{The reconstruction behavior of RCD on the entire Yoochoose dataset (including train, val, and test).
    }
    \label{fig: cvae_beta_tunning}
\end{figure*}

\begin{figure}[t]
    \begin{center}
        \includegraphics[width=0.48\textwidth]{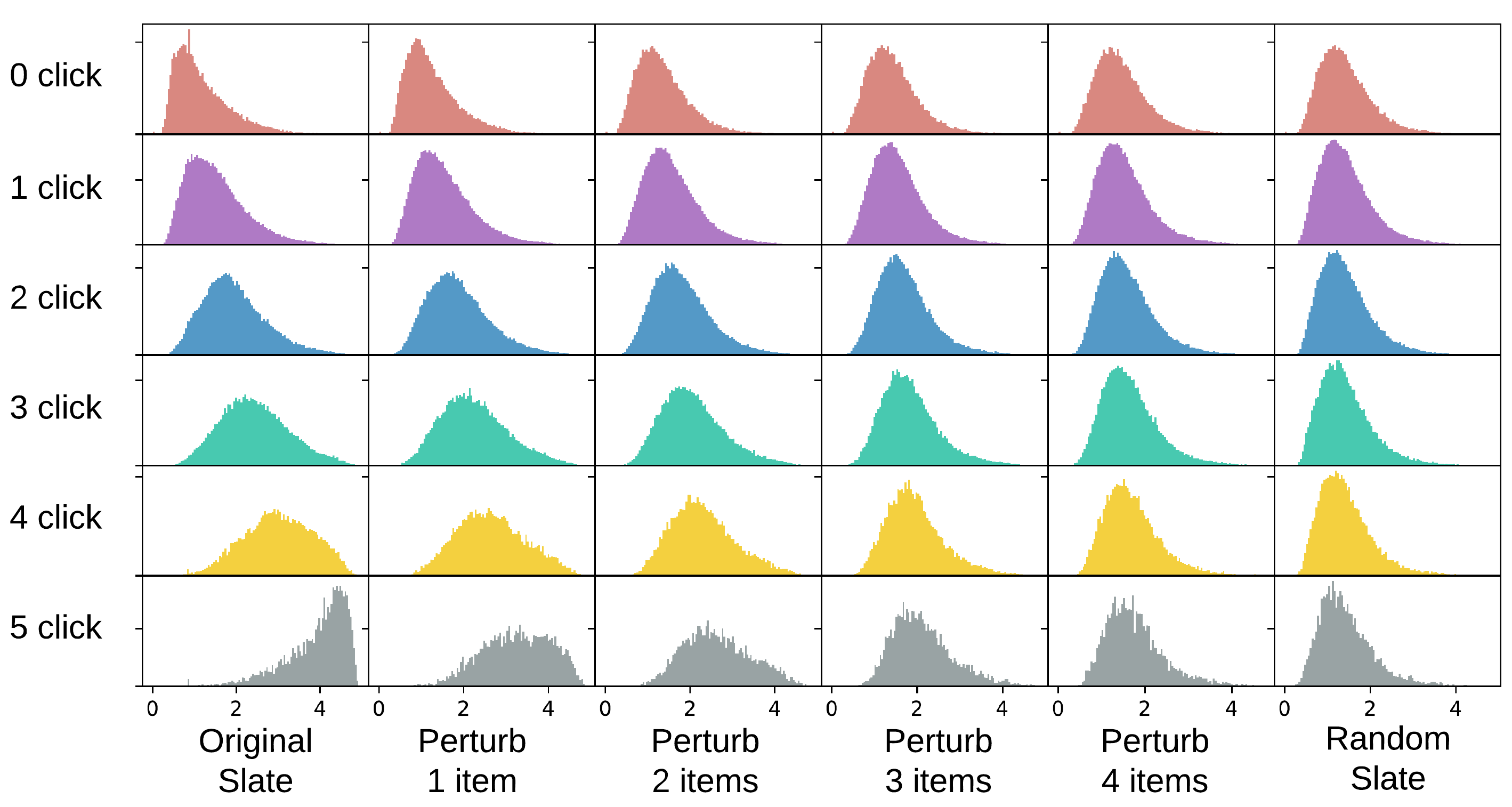}
        \caption{User response distribution under perturbation.
        }
        \label{fig: change_one_item}
    \end{center}
% \end{wrapfigure}
\end{figure}

\section{Post-Generation Perturbation}\label{ap: perturbation}

Given a slate (of size 5) and its observed labels (\#click) in data, we choose $a$ numbers of its items randomly and apply perturbation to observe how much the ground-truth user response distribution $R(r|\bm{s})$ is affected.
We present the result of the entire Yoochoose data (including val and test) in Figure \ref{fig: change_one_item}.
% Each subplot represents the user response distribution of slates with the same observed label and the same perturbation number $a$.
Each row corresponds to slates with certain observed label (\#click), and each column corresponds to the number of perturbed item $a$.
In each subplot, the $x$-axis corresponds to the ground truth expected user response $R(r|\bm{s})$ (from pre-trained model if Yoochoose/MovieLens dataset, and from simulation if URM-based environments), and the $y$-axis corresponds to the frequency/density of slates.
As shown in the first column where no perturbation is involved, $R(r|\bm{s})$ is usually very close to the observed label.
However, as given by the second column from left, changing merely a single item has already caused a significant deviation of distribution from that of the original slates, especially for slates with ideal condition $\bm{r}^*$ (the bottom row).
And perturbation of 3 items is already inducing a distribution close to that of random slates (the rightmost column).
We also observed this substantial reduction in MovieLens and simulation data.
As shown in table \ref{tab: full}, simply applying post-generation perturbation on a given slate without considering the context of the slate neither achieves the best ENC nor the best variation.

\section{More on RCD}\label{ap: rcd}

The detailed view of the reconstruction behavior of List-CVAE on the entire Yoochoose data is given by Figure \ref{fig: cvae_beta_tunning}.
The same pattern also appears on MovieLens 100K.
Each subplot gives the result of a List-CVAE model with certain $\beta$, the $y$-axis represents the predicted user response $R(\bm{r}|s)$ of the original slate and the $x$-axis represents $R(\bm{r}|\widehat{s})$ of the reconstructed slates.
The reconstruction behavior on the dataset reveals its performance of inferring the observed dataset (including test set), which also helps identifying the RCD.
Slates with the same observed \#click have the same color, so ideally the slates with the same color should be somewhere close to the location [\#click, \#click].
We observe that the over-reconstruction happens when $\beta \leq 0.003$ (first three subplots), and the more distinguishable diagonal line in the plot with $\beta=0.0003$ indicates a more severe reconstruction overfitting.
The over-concentration happens when $\beta > 0.015$, and the reconstructed slates are typically gathered to the very point where the prototype slates are.
This means that the decoder always tends to generate the same slate based only on the given constraint.
Finally, the ``elbow'' transition between these two extreme cases happens in a narrow region around $0.01$, where we observe several ``prototypes'' (the clustered vertical lines of each color) in each response group.
In this region, the model shows moderate concentration behavior, but the resulting slates may still cover some degree of variance.

\section{Slate Response Distribution}
\label{sec:slate_res}

\begin{figure}[t]
    \centering
    \includegraphics[width=0.46\textwidth]{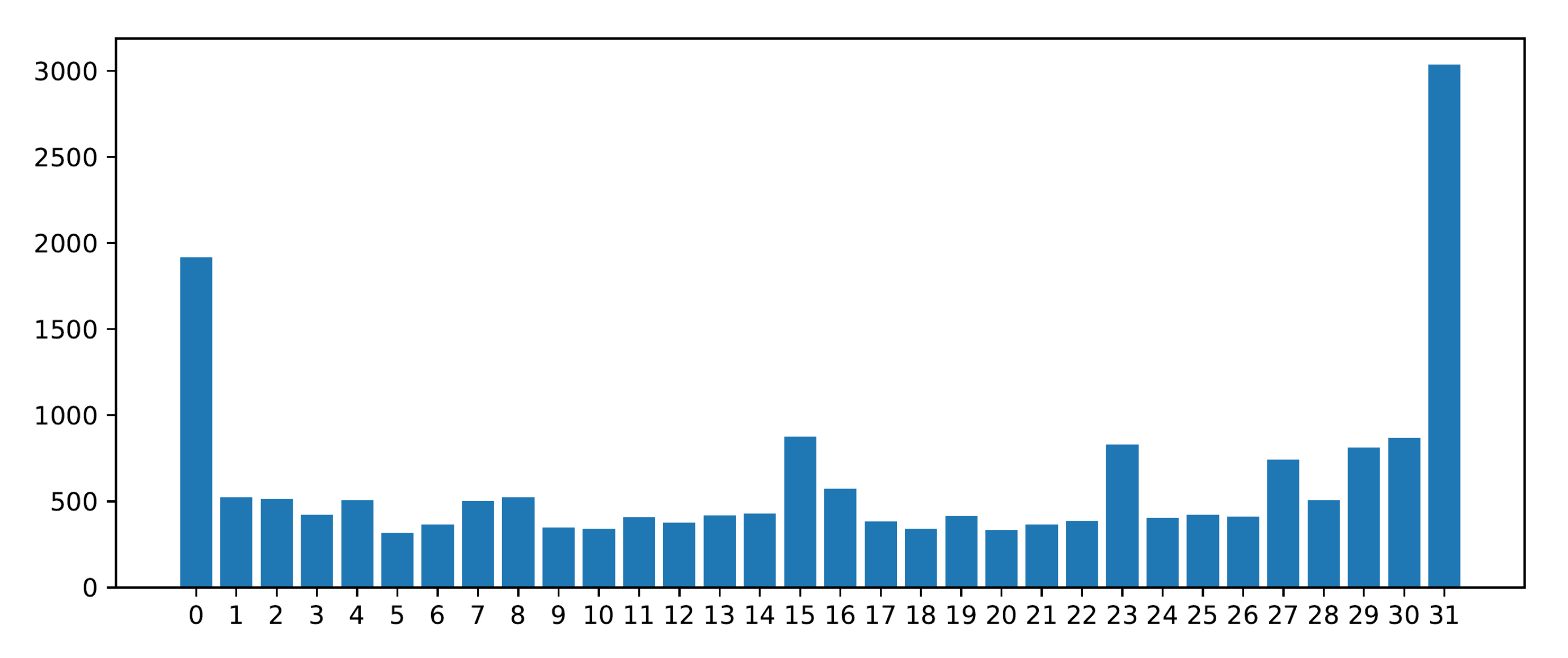}
    \caption{
    Slate response distribution of the preprocessed MovieLens 100K dataset with slate size $k=5$.
    % {\color{red}The periodical spike reveals the existence of positional bias.}
    }
    \label{fig: slate_resp_dist}
\end{figure}

The resulting slate response distribution of Yoochoose data and MovieLens 100K data are similar, and we show the later in Figure \ref{fig: slate_resp_dist}.
Y-axis represents the frequency and X-axis correspond to the ground truth response $\bm{r}$ of slates.
The label of X-axis is obtained by considering each $\bm{r}$ as a binary number and expressing it as integer.
For example, user response $[0,0,0,0,1] \rightarrow 1_2 \rightarrow 1_{10}$ and $[1,1,1,1,0] \rightarrow 11110_2 \rightarrow 30_{10}$ where the subscript 2 and 10 means binary and decimal representation of numbers.

\section{Design of Simulation}\label{ap: simulation}

\textbf{Basic User Response Model (URM):} 
We assume that the basic interaction between users and items follow a user interest model, which is a biased matrix factorization model\cite{koren2009matrix}.
Each item $d^i \in \mathcal{D}$ is associated with a vector $\bm{v}_i \in \mathbb{R}^m$, where $m$ is the embedding dimension.
Each user $j$ is assigned with a vector of interest $\bm{u}_j \in \mathbb{R}^m$. 
To find realistic settings, we first observe the distribution of user embeddings, item embeddings, user biases, and item biases in pre-trained $R(\bm{r}|\bm{s})$ of MovieLens dataset, then use the same mean and variance to randomly sample each $\bm{v}_i$, $\bm{u}_j$, user bias $b^u_j$, item bias $b^v_j$, and global bias $b$.
And the user's initial interest in $d^i$ is given by:
\begin{equation*}
    \mathcal{I}_{\mathrm{URM}}(d^i, j) = g(\bm{u}_j^T \bm{v}_i + b_j^u + b_i^v + b)
\end{equation*}
where $g$ is a Sigmoid function.
This basic model assumes independent point-wise interaction for each user-item pair and no other effect from the slate context.

\textbf{Adding Positional Bias (URM\_P):} Items appeared at the previous positions of the slate are assumed to have a higher chance of receiving positive feedback than those in later positions.
The reason for this design is that users may gradually lose their patience when further browsing the items~\cite{ie2019slateq}.
In our setting, we first employ an average positional offset $\bm{b}^p=[0.2,0.1,0.0,-0.1,-0.2]$.
And for each user, we draw personalized positional bias $\mathcal{B}(j) \sim \mathcal{N}(\bm{b}^p, \bm{\sigma}_u^2)$ where the variance $\bm{\sigma}_u^2 = 0.2$.
Then the final probability of click:
\begin{equation*}
    \mathcal{I}_{\mathrm{URM\_P}}(d^i,j) = \mathrm{clip}\big(\mathcal{I}_{\mathrm{URM}}(d^i,j) + \lambda \mathcal{B}(j)_k, 0, 1)\big)
\end{equation*}
where $\lambda$ (set to 1.0 during experiment) controls how significant is the impact of positional bias on the user responses.
The clip function ensures that the user's interest is within $[0,1]$.

\textbf{Adding Item Interactions (URM\_P\_MR): } 
In \cite{jiang2018beyond}, the authors assumed that item interactions are combinations of binary relations.
Here we use a simple and easy-to-control multi-item relation model: 
First assume that a user's attention is altered when she sees the overall features of the slate:
\begin{equation*}
    \mathrm{Atn}(\bm{s},j) = g\bigg(\Bigl(\frac{1}{K}\sum_{d^i \in s}(\bm{v}_i)\Bigr) \odot \bm{u}_j\bigg)
\end{equation*}
where $\odot$ denotes element-wise multiplication.
Then the resulting attention is applied to each item to obtain the excursion:
\begin{equation*}
    \mathcal{M}(\bm{s},j)_i = \mathrm{Atn}(\bm{s},j)^T \bm{v}_i
\end{equation*}
Add up everything so far gives the final probability of click:
\begin{equation*}
     \mathcal{I}_{\mathrm{URM\_P\_MR}}(d^i,j) =  \mathrm{clip}\big(\mathcal{I}_{\mathrm{URM}}(d^i,j) + \lambda \mathcal{B}(j)_k + \rho \mathcal{M}(\bm{s},j)_i, 0, 1\big)
\end{equation*}
where coefficient $\rho$ is introduced to control the significance of the item relation term.
Though we found these simulation settings sufficient for our study, one may sue for more realistic and advanced simulation like~\cite{ie2019recsim} as complementary approaches.

\subsection{Stochastic vs. Deterministic}\label{sec: gen_vs_dis}

Compare to generative models, discriminative ranking models are deterministic and cannot explore the variety of slates, thus they are favored by ranking metrics on offline test set. 
We present the ranking performance on test set in Table \ref{tab: ranking}-D and \ref{tab: ranking}-E.
For most datasets, different from the ``ground truth'' user response evaluated by environment $R(r|\bm{S})$, generative models (CVAE-based models) are stochastic and tend to explore more choices of good but unseen slates beyond the limited observation of the test set.
On the other hand, ranking models like MF and NeuMF tend to focus on the best point that satisfies users the most, and perturbation of just one item does not severely harm the ranking metrics of the whole slate since the remaining items are also individually accurate.
Interestingly, the gap between deterministic and stochastic models becomes less observable when we increase the proportion of item relations in the slates (URM\_P $\rightarrow$ URM\_MR), and generative models even start to outperform MF and NeuMF when $\rho = 5.0$.
This shows that generative models are able to model whole-slate patterns for which MF and NeuMF often fail to learn.

\begin{table}[t]
    \centering
    \caption{Infeasible Evaluation on Test Set, see section \ref{sec: gen_vs_dis}}
    \begin{adjustbox}{max width=\linewidth}
    \begin{tabular}{ccccc}
    \toprule
         & ML(User) & URM\_P & $\substack{\textrm{URM\_P\_MR}
         \\(\rho=0.5)}$ & $\substack{\textrm{URM\_P\_MR}\\(\rho=5.0)}$\\
        \bottomrule
        \toprule
        \multicolumn{5}{c}{D: Slate Hit Rate}\\
        \midrule
        MF & \textbf{0.0330} & 0.0069 & 0.0071 & 0.0092\\
        NeuMF & 0.0272 & \textbf{0.0089} & \textbf{0.0088} & 0.0068\\
        MF-MMR & 0.0092 & 0.0057 & 0.0078 & 0.0070 \\
        \midrule
        Non-Greedy MF & 0.0283 & 0.0063 & 0.0072 & 0.0087\\
        Non-Greedy NeuMF & 0.0233 & 0.0077 & 0.0080 & 0.0072 \\
        \midrule
        List-CVAE & 0.0041 & \underline{0.0071} & \underline{0.0079} & \underline{\textbf{0.0093}} \\
        \midrule
        Non-Greedy List-CVAE & 0.0056 & 0.0068 & 0.0078 & 0.0090\\
        Pivot-CVAE (SGT-PI) & 0.0043 & \underline{0.0071} & 0.0069 & 0.0078\\
        Pivot-CVAE (GT-SPI) & \underline{0.0131} & 0.0062 & 0.0072 & 0.0080\\
        Pivot-CVAE (SGT-SPI) & 0.0043 & 0.0070 & 0.0072 & 0.0074\\
        \bottomrule
        \toprule
        \multicolumn{5}{c}{E: Slate Recall}\\
        \midrule
        MF & \textbf{0.0090} & 0.0021 & 0.0024 & 0.0034\\
        NeuMF & 0.0079 & \textbf{0.0029} & \textbf{0.0029} & 0.0024\\
        MF-MMR & 0.0032 & 0.0018 & 0.0026 & 0.0025 \\
        \midrule
        Non-Greedy MF & 0.0078 & 0.0019 & 0.0025 & 0.0032\\
        Non-Greedy NeuMF & 0.0072 & 0.0025 & 0.0026 & 0.0026\\
        \midrule
        List-CVAE & 0.0013 & \underline{0.0024} & \underline{\textbf{0.0029}} & \underline{\textbf{0.0038}} \\
        \midrule
        Non-Greedy List-CVAE & 0.0021 & 0.0022 & 0.0027 & 0.0035\\
        Pivot-CVAE (SGT-PI) & 0.0014 & 0.0023 & 0.0024 & 0.0026\\
        Pivot-CVAE (GT-SPI) & \underline{0.0038} & 0.0020 & 0.0024 & 0.0028\\
        Pivot-CVAE (SGT-SPI) & 0.0013 & 0.0023 & 0.0025 & 0.0028\\
        \bottomrule
    \end{tabular}
    \end{adjustbox}
    \label{tab: ranking}
\end{table}

\end{document}